\documentclass[aps,prl,twocolumn,amsfonts,amssymb,amsmath,floats,floatfix,showpacs,preprintnumbers,superscriptaddress]{revtex4}

\usepackage{graphicx}
\usepackage{hyperref}

\hypersetup{
colorlinks=true,
urlcolor= blue,
citecolor=blue,
linkcolor= blue,
bookmarks=true,
bookmarksopen=false,
}
\usepackage[pdftex]{thumbpdf}
\usepackage{graphicx}
\usepackage{dcolumn}
\usepackage{bm}
\usepackage{color}
\usepackage{multirow}
\usepackage{amsmath}

\newcommand{\SR}{Sr$_2$RuO$_4$}
\newcommand{\dg}{$^{\circ}$}
\newcommand{\rtr}{$(\!\sqrt{2}\!\times\!\!\sqrt{2})$R45\dg{}}

\begin{document}

\title{Determining the surface-to-bulk progression in the normal-state electronic structure \\ of
\SR{} by angle-resolved photoemission and density functional theory}

\author{C.\,N. Veenstra}
\author{Z.\,-H. Zhu}
\author{B.\, Ludbrook}
\author{M. Capsoni}
\affiliation{Department of Physics {\rm {\&}} Astronomy, University of British Columbia, Vancouver, British Columbia V6T\,1Z1, Canada}
\author{G.\, Levy}
\author{A.\, Nicolaou}
\affiliation{Department of Physics {\rm {\&}} Astronomy, University of British Columbia, Vancouver, British Columbia V6T\,1Z1, Canada}
\affiliation{Quantum Matter Institute, University of British Columbia, Vancouver, British Columbia V6T\,1Z4, Canada}
\author{\\J.\,A. Rosen}
\author{R.\, Comin}
\affiliation{Department of Physics {\rm {\&}} Astronomy, University of British Columbia, Vancouver, British Columbia V6T\,1Z1, Canada}
\author{S. Kittaka}
\author{Y. Maeno}
\affiliation{Department of Physics, Graduate School of Science, Kyoto University, Kyoto 606-8502, Japan}
\author{I.\,S. Elfimov}
\author{A.\, Damascelli}
\email{damascelli@physics.ubc.ca}
\affiliation{Department of Physics {\rm {\&}} Astronomy, University of British Columbia, Vancouver, British Columbia V6T\,1Z1, Canada}
\affiliation{Quantum Matter Institute, University of British Columbia, Vancouver, British Columbia V6T\,1Z4, Canada}

\date{\today}

\begin{abstract}
We revisit the normal-state electronic structure of \SR{} by angle-resolved photoemission spectroscopy (ARPES) with improved data quality, as well as ab-initio band structure calculations in the local-density approximation (LDA) with the inclusion of spin-orbit coupling (SO). We find that the current model of a single surface layer $(\sqrt{2} \times \sqrt{2})$R45\dg{} reconstruction does not explain all detected features. The observed depth-dependent signal degradation, together with the close quantitative agreement with LDA+SO slab calculations based on the surface crystal structure as determined by low-energy electron diffraction (LEED), reveal that -- at a minimum -- the subsurface layer also undergoes a similar although weaker reconstruction. This model accounts for all features -- a key step in understanding the electronic structure - and indicates a surface-to-bulk progression of the electronic states driven by structural instabilities, with no evidence for other phases stemming from either topological bulk properties or the interplay between SO and the broken symmetry of the surface.
\end{abstract}
 
\pacs{74.25.Jb, 74.70.Ad, 79.60.Jv, 79.60.Bm}

\maketitle

Since its discovery \cite{Nature.372.532}, understanding the origin of superconductivity in \SR{} has received enormous attention \cite{JPSJ.81.011009}. In search of the prodromes of the elusive pairing mechanism, its normal-state electronic structure has also been a matter of intense study. A unified picture of the Fermi surface (FS) was obtained from bulk-sensitive de Haas-van Alphen (dHvA) measurements \cite{PhysRevLett.76.3786,PhysRevLett.84.2662} and surface sensitive angle-resolved photoemission spectroscopy (ARPES) \cite{PhysRevLett.85.5194}. This, together with local-density approximation (LDA) band structure calculations \cite{PhysRevLett.101.026406,PhysRevLett.101.026408}, established the importance of spin-orbit coupling (SO) in defining the low-energy electronic structure. The inclusion of SO may also lie at the root of the bigger question regarding the superconducting mechanism, with suggestions that it may facilitate mixed-parity pairing \cite{PhysRevLett.101.026406,PhysRevLett.107.277003,arXiv1101.4656}. 

Similar to the case of topological insulators the presence of SO also raises the possibility that novel topological states, Dirac or Rashba-type, might exist on the (001) cleaving-surface of \SR{} in the normal-state (in addition to the chiral superconductivity topological edge-states detected below $T_c$ on the side surfaces  \cite{PhysRevLett.107.077003}). This proposal is particularly intriguing given that surface-specific electronic states have been observed in this material  \cite{PhysRevLett.85.5194,PhysRevB.64.180502}. 
\begin{figure}[br!]
\includegraphics[width=1.0\linewidth]{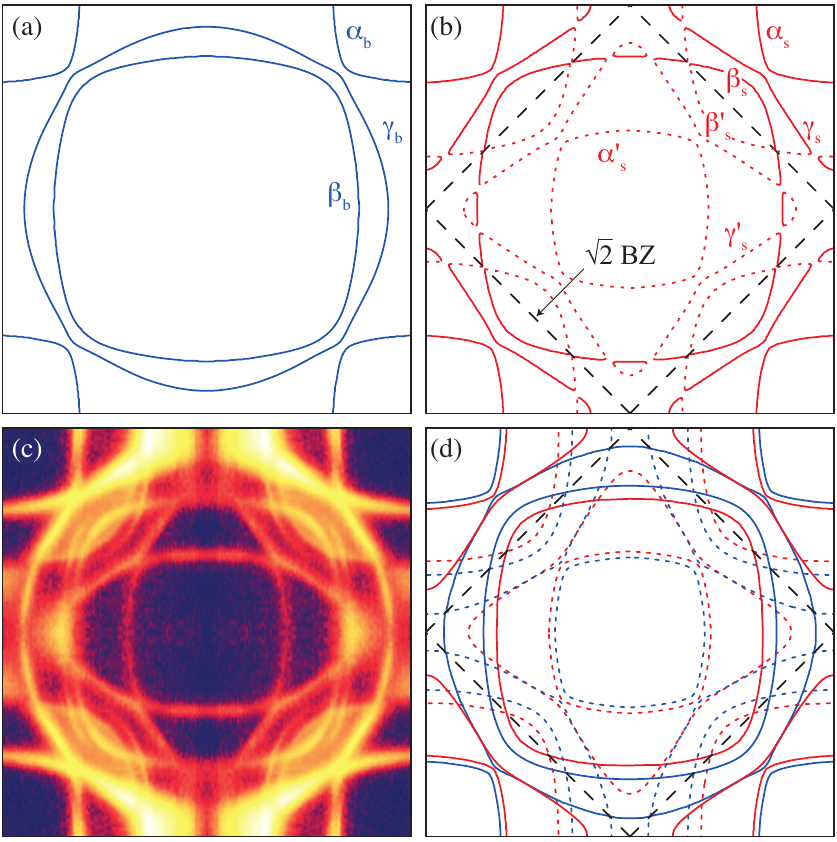}
\caption{(color online). The LDA+SO, $k_z\!=\!0$ FS of \SR{} for: (a) undistorted bulk; (b) bulk with RuO$_6$ octahedra rotated 5.5\dg{} causing a \rtr{} reconstruction (new BZ - dashed; unfolded FS - solid; folded replica FS - dotted). (c) ARPES FS (7\,meV integration) from a fresh low-temperature cleave; (d) phenomenological FS based on (c), Fig.\,\ref{splitting}, and \ref{time}.
}\label{FS}
\end{figure}
\begin{figure*}
\includegraphics[width=1.0\linewidth]{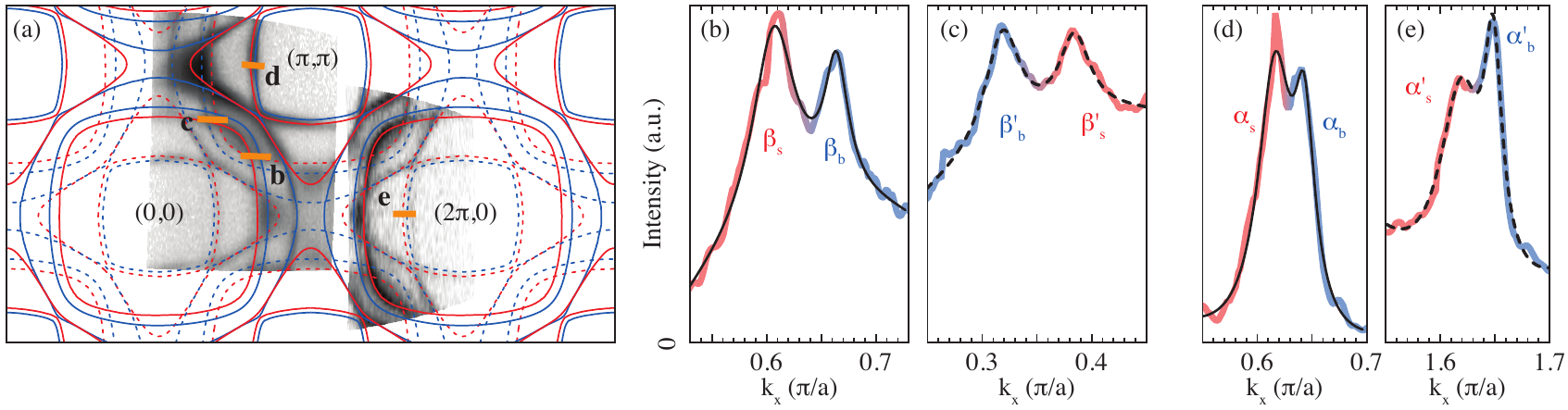}
\caption{(color online). (a) \SR{} FS maps (grey scale), marking the location of MDC cuts at the Fermi energy showing clear doubling in the (b) $\beta$ sheet, (c) its folded replica $\beta^\prime$, (d) $\alpha$, and (e) folded $\alpha^\prime$.  (b-e) Data are shown in color, with red and blue referring to surface and bulk-like features, while fit results (see text) are shown in black (dashed lines for folded bands).}\label{splitting}
\end{figure*}
While these have originally been interpreted in terms of a purely structural reconstruction of the top surface layer  \cite{PhysRevLett.85.5194,PhysRevB.64.180502} as resolved via low-energy electron diffraction (LEED) \cite{Science.289.746}, a recent ARPES study detected an additional dispersive band \cite{2011arXiv1103.6196Z}, demanding this simple model be revisited in terms of a possible manifestation of topological bulk properties or Rashba interaction.

Here we study the normal-state electronic structure of \SR{} by ARPES and density-functional theory (DFT).
With improved data quality a multitude of new dispersive features can be detected, stemming from the folding of \emph{all} FSs: reconstructed-surface's and bulk-like's as well. While this might be suggestive of surface ferromagnetism or Rashba-type splitting, we show -- through controlled aging and LDA+SO slab calculations for the LEED-determined crystal structure \cite{MooreThesis} -- that this doubling arises from surface and subsurface instabilities. This discovery bears close similarity to the recent findings of surface-enhanced charge-density-wave \cite{arXiv1111.2673} and stripe \cite{NatComm.3.1023} order in the underdoped high-T$_{\text{c}}$ cuprates, establishing structural/electronic  instabilities in the near surface region a widespread phenomenon of complex oxides. 

\begin{figure}
\includegraphics[width=1.0\linewidth]{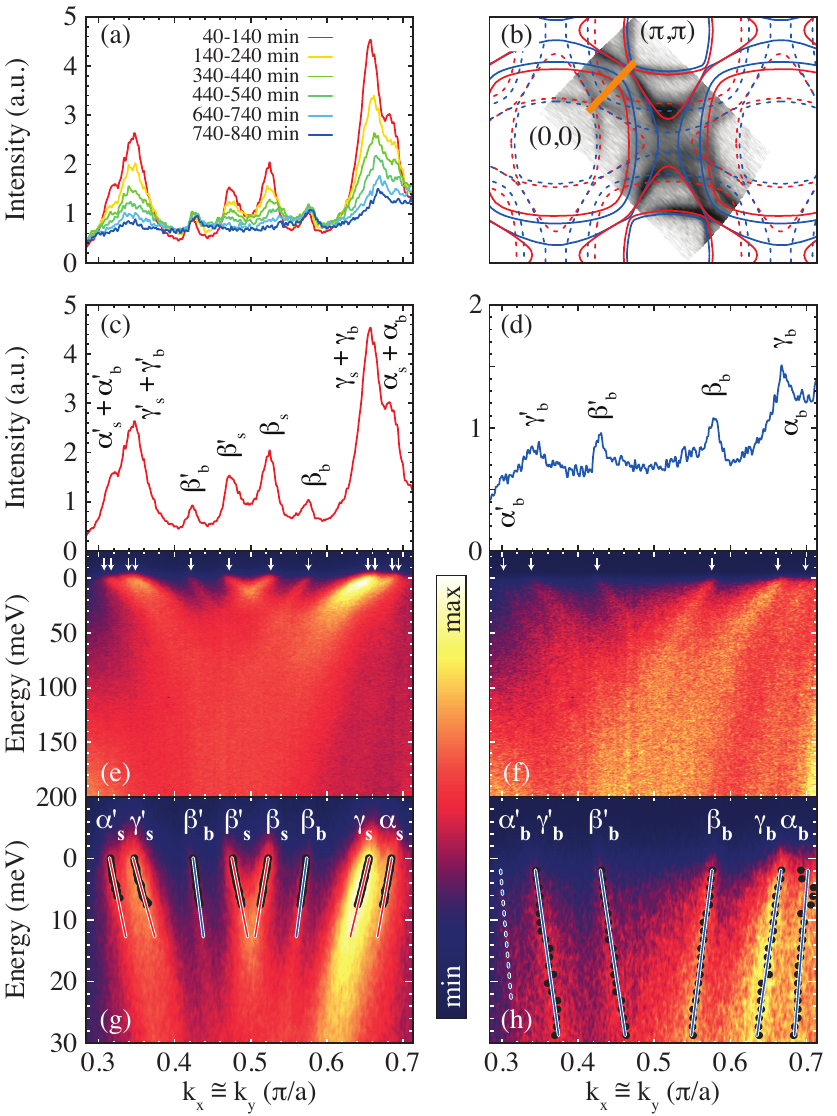}
\caption{(color online). (a) Time evolution of a 45$^\circ$ MDC cut at $E_F$; its location is marked on the alignment FS in (b). (c,e,g) Plot of the corresponding MDC at $E_F$, band map of features on a large energy range, and zoom in energy on the Fermi crossings for the first 100\,minutes after cleave; (d,f,h) the same for the final 100\,minutes of acquisition. In (g,h), MDC maxima (black circles) and MDC-dispersion fits (blue-bulk/red-surface lines) are also shown.}\label{time}
\end{figure}

ARPES experiments were performed at UBC with a SPECS Phoibos 150 analyzer and 21.2\,eV photons from a monochromatized UVS300 lamp. FWHM energy and momentum resolutions were measured to be 17\,meV and 0.01\,$\frac{\pi}{a}$ angular equivalent in Fig.\,\ref{FS} and \ref{splitting}, and 9\,meV and 0.04\,$\frac{\pi}{\sqrt{2}a}$ in Fig.\,\ref{time}. Samples grown by the floating-zone method \cite{JPSJ.68.1651} were oriented by Laue diffraction, cleaved in-situ, then held at temperatures between 5-10\,K and pressures varied in the $5\!-\!9\!\times\!10^{-11}$\,mbar range. To verify that the observed ARPES features are not associated with Ru-metal inclusions, the so-called `3K-phase'  \cite{JPSJ.68.1651}, samples with and without inclusions were studied: no 3K-phase dependence was observed.

\begin{figure}[b]
\includegraphics[width=1.0\linewidth]{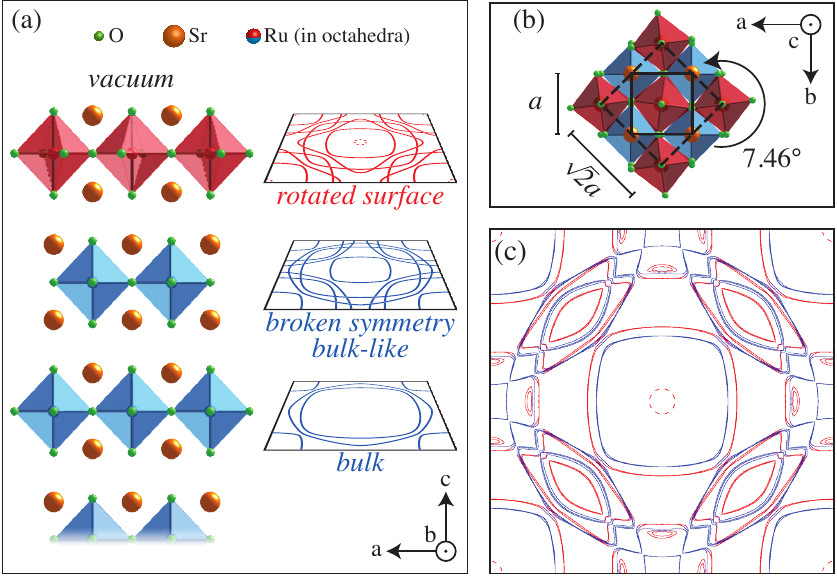}
\caption{(color online). (a) Partial structure of the \SR{} 5-layer slab used for LDA+SO calculations, with corresponding surface and bulk-like FS. (b) Reconstructed surface of \SR{}, with original undistorted unit cell shown solid and new shown dashed. (c) FS obtained from the LDA+SO slab calculations by shifting the chemical potential 24\,meV down in energy to match the slight averaged hole-doping observed by ARPES for \SR{} surface and bulk-like FS. Surface and bulk character are shown in red and blue, respectively.}\label{ldaslab}
\end{figure}

Let us begin with a brief review of the previously accepted electronic structure from bulk and reconstructed surface  \cite{PhysRevLett.85.5194,PhysRevB.64.180502,PhysRevLett.101.026406}. Fig.\,\ref{FS}(a) shows the LDA+SO bulk FS derived from the Ru-$t_{2g}$ orbitals occupied by 4 electrons, which is in good agreement with results from dHvA  \cite{PhysRevLett.76.3786,PhysRevLett.84.2662}, as well as ARPES on an optimally degraded surface \cite{PhysRevLett.85.5194}. As a consequence of the \rtr{} surface reconstruction observed by LEED \cite{Science.289.746} and associated with the rotation of the RuO$_6$ octahedra [Fig.\,\ref{ldaslab}(c)], the Brillouin Zone (BZ) is reduced leading to band folding; this is shown in Fig.\,\ref{FS}(b) for a 5.5\dg{}-rotation LDA+SO bulk calculation (note that primes and dotted lines will be used for folded FSs, and $s$/red for surface and $b$/blue for bulk-like states). In Fig.\,\ref{FS}(c) we present the ARPES FS for a pristine cleave of \SR{}, measured in $p$-polarization at 5\,K and $6\!\times\!10^{-11}$\,mbar in less than 3 hours after cleaving, during which time no aging was observed. The data were acquired over the entire upper right quadrant without symmetrization, then folded in $k$-space to reproduce the BZ as shown (the analyzer entrance slit was oriented horizontally with respect to the final image, giving greater angular resolution in that direction). One can recognize doubling and folding of many features, seemingly consistent with the overlap of bulk and reconstructed FSs from Fig.\,\ref{FS}(a) and (b), as originally proposed \cite{PhysRevLett.85.5194,PhysRevB.64.180502}. For instance, both $\beta_b$ and $\beta_s$ are seen, as well as both $\gamma_b$ and $\gamma_s$ with their expected electron and hole-like topology [with $\gamma_s$ showing clearly near $(0,\pi)$ and $\gamma_b$ near $(\pi,0)$, due to varying matrix element dependence]. However, contrary to this explanation, the folded $\beta^\prime$ also is clearly doubled giving rise to $\beta_b^\prime$ and $\beta_s^\prime$. We will show in Fig.\,\ref{splitting} and \ref{time} that in fact \emph{all} bands are doubled --  both unfolded \emph{and folded} -- resulting in the phenomenological FS of Fig.\,\ref{FS}(d). This cannot be explained by a single reconstructed surface-layer model, and may instead be consistent with the interesting proposal of SO-driven surface effects \cite{2011arXiv1103.6196Z}.

In Fig.\,\ref{splitting} we confirm the doubling of features with momentum distribution curves (MDCs) obtained by integrating a 10\,meV region just above $E_F$: panels (b-e) show the doubling of $\beta$, $\beta^\prime$, $\alpha$, and $\alpha^\prime$, respectively, observed at the $k$-locations indicated in the alignment FS of (a). Note that the splitting between $\alpha^\prime_s$ and $\alpha^\prime_b$ is more clearly resolved in the second BZ [Fig.\,\ref{splitting}(e)], perhaps due to matrix element effects and/or $2^{nd}$-BZ higher angular resolution. By performing a least-squares fit to two Lorentzian peaks with a second-order polynomial background (black curves), we can estimate the observed splittings as: 0.057, 0.065, 0.027, and 0.029$\pm$0.003\,$\frac{\pi}{a}$ for $\beta$, $\beta^\prime$, $\alpha$, and $\alpha^\prime$. Although it is difficult to compare these values directly, due to the slight difference in angles and $k$-space locations between them, we note that the splittings between $\beta_s/\beta_b$ and $\beta^\prime_s/\beta^\prime_b$ are very close and approximately double those between $\alpha_s/\alpha_b$ and $\alpha^\prime_s/\alpha^\prime_b$, which are almost identical.

Thus far, with a doubling observed for all $\alpha$ and $\beta$ bands and in light of the bulk magnetism of many Ru-oxides, one might be drawn to the possibility of Rashba-type effects \cite{2011arXiv1103.6196Z} or even surface ferromagnetism \cite{Science.289.746}. If any of these were the case one would expect that -- under surface degradation -- the surface states' splitting would shrink and/or the intensity decay uniformly over time for all of them, eventually revealing the underlying bulk electronic structure. In Fig.\,\ref{time} we show the results of such a time-dependent experiment for the band dispersion along the $(0,0)\!-\!(\pi,\pi)$ direction in $s$-polarization, where all features and their folded replicas are visible simultaneously, for a sample maintained at 9.5\,K while raising the base pressure to about 8.5$\times10^{-11}$\,mbar. 
\begin{table*}[t!]
  \begin{tabular}[t]{ | p{1.6cm}<{\centering} *{2}{ | p{2.6cm}<{\centering} p{2.6cm}<{\centering}   } | c | p{1.7cm}<{\centering} | }
\hline
   Band &   $n_e^\text{LDA+SO}$ & $v_F^\text{LDA+SO}$ (eV \AA{}) & $n_e^\text{ARPES}$ & $v_F^\text{ARPES}$ (eV \AA{}) & $v_F^\text{LDA+SO}/v_F^\text{ARPES}$ & $n_e^\text{dHvA}$ \\
  \hline
  $\alpha_s, \alpha_s^\prime$      & 1.668 $\pm$ 0.007 & 2.30 $\pm$ 0.01 & 1.721 $\pm$ 0.040 & 0.56 $\pm$ 0.05 & 4.1  $\pm$  0.4 & -- \\
  $\beta_s, \beta_s^\prime$          & 0.864 $\pm$ 0.006 & 2.14 $\pm$ 0.01 & 0.757 $\pm$ 0.020 & 0.59 $\pm$ 0.05 & 3.6  $\pm$  0.3 & -- \\
  $\gamma_s, \gamma_s^\prime$ & 1.340 $\pm$ 0.011 & 2.18 $\pm$ 0.02 & 1.396 $\pm$ 0.060 & 0.42 $\pm$ 0.05 & 5.2  $\pm$  0.6 & -- \\

total & 3.872 $\pm$ 0.014 & -- & 3.874 $\pm$ 0.120 & -- & -- & -- \\  
  
  \hline
 $\alpha_b, \alpha_b^\prime$       & 1.740 $\pm$ 0.003 & 2.14 $\pm$ 0.02 & 1.760 $\pm$ 0.040 & 1.30 $\pm$ 0.60 & 1.6  $\pm$  0.8 & 1.781 \\

  $\beta_b, \beta_b^\prime$  & 0.967 $\pm$ 0.022 & 2.60 $\pm$ 0.25 & 0.903 $\pm$ 0.020 & 0.80 $\pm$ 0.10 & 3.3  $\pm$  0.5 & 0.921 \\
  
  $\gamma_b, \gamma_b^\prime$ & 1.252 $\pm$ 0.014 & 2.21 $\pm$ 0.07 & 1.280 $\pm$ 0.060 & 0.76 $\pm$ 0.10 & 2.9  $\pm$  0.4 & 1.346 \\
  
    total & 3.959 $\pm$ 0.026 & -- & 3.943 $\pm$ 0.120  & -- & -- & 4.048 \\
\hline
\end{tabular}
\caption{Carrier concentrations counting electrons, $n_e\!=\!2\!\times\!A_\text{FS}/A_\text{BZ}$ with 2 accounting for the spin degeneracy, and Fermi velocities along the $(0,0)-(\pi,\pi)$ direction, $v_F$, as determined by our ARPES and LDA+SO slab calculations for surface and bulk-like electronic structure. ARPES-FS volume estimates are from the phenomenological FS in Fig.\,\ref{FS}; LDA+SO results were obtained for the 24\,meV shifted chemical potential to match the average electron counting of surface and subsurface FSs as determined by ARPES. The dHvA results from Ref.\,\onlinecite{PhysRevLett.84.2662}, representative of the bulk Luttinger's counting, are also shown.}
\label{table}
\end{table*}

The time-evolution of MDCs integrated over 100 minutes and 4\,meV just above $E_F$ is presented in Fig.\,\ref{time}(a): while some features remain at a similar level of intensity  over the 14-hour time-span, others age and have their intensity suppressed, without any overall change in slope or shape. Particularly comparing Fig.\,\ref{time}(c,e,g) and (d,f,g), which present only first and final MDC and dispersion results, we can infer that: (i) \emph{all} bands and their folded replicas are doubled for a total of 12 bands (as labelled), including the $\gamma$ bands not previously demonstrated in Fig.\,\ref{splitting}; (ii) based on the observed difference in degradation rate, the electronic states do not all originate from the very same surface layer, and can instead be classified as surface (fast degradation) or bulk-like (slow degradation). 
These observations can be made more quantitative from the MDC-dispersion fit analysis, which allows distinguishing between surface and bulk-like bands based on their remarkably different Fermi velocity $v_F$ (see Tab.\,\ref{table}). While the fresh data in Fig.\,\ref{time}(g) are dominated by the surface  $\alpha_s$/$\alpha^\prime_s$, $\beta_s$/$\beta^\prime_s$, and $\gamma_s$/$\gamma^\prime_s$, with the addition of the well-separated $\beta_b$/$\beta^\prime_b$ pair, the aged data of Fig.\,\ref{time}(h) only show the much steeper bulk-like $\alpha_b$/$\alpha^\prime_b$, $\beta_b$/$\beta^\prime_b$, and $\gamma_b$/$\gamma^\prime_b$ ($\alpha^\prime_b$, although visible, could not be fit and is shown with a dashed line as a guide to the eye). 

Since all $s$-states are suppressed evenly and independently from the $b$-ones, we must conclude that neither ferromagnetism or Rashba coupling is responsible for the apparent band doubling. Similarly, we must rule out possible surface patches with different structure or effective doping, as well as bulk photoelectrons being scattered off the reconstructed surface 
\cite{FN1}.
As intrinsic bulk properties can also be excluded, otherwise they would be visible in dHvA \cite{PhysRevLett.76.3786,PhysRevLett.84.2662} and x-ray diffraction \cite{ActaCrystC.49.1268}, the most natural explanation would be that the subsurface layer has a structural modulation similar to, yet weaker, than that found at the surface. This lesser modulation would be enough to break the symmetry of the BZ and cause a visible folding, but not enough to strongly modify FS topologies and volumes. The subsurface layer is then responsible for all $b$-states while the surface for $s$-states. Measured FS volumes [Fig.\,\ref{FS}(c,d)] support this model: as shown in Tab.\,\ref{table}, while the $s$-FS electron counting is somewhat reduced ($n_e^s\!\simeq\!3.87$), the $b$-FS electron counting is much closer to the dHvA results adding up to $n_e^b\!\simeq\!3.94$, i.e. almost 4 electrons as expected in the Ru $t_{2g}$ orbitals (this still seems to suggest the possibility that these near-surface layers are slightly hole-doped, possibly due to the cleaving-induced Sr-vacancies revealed by scanning tunnelling microscopy \cite{yan}). Finally photoemission from a second layer, also folded but otherwise similar to the bulk, could even explain the so-called `shadow band' which remained after high-temperature cleaving had suppressed the surface states in the original work uniting the dHvA and ARPES pictures of the FS of \SR{} \cite{PhysRevLett.85.5194}.

To conclusively validate this explanation we have performed a 5-RuO$_2$-layer LDA+SO slab calculation using the linearized augmented-plane-wave method in the WIEN2K package \cite{wein2k}.  Explicitly including these layers and the vacuum in the unit cell allows us to study both the structural reconstruction as well as effects due to broken symmetry in proximity of the surface in a self-consistent way.  Additionally, this enables us to estimate the surface/bulk character of the resulting bands by projecting the corresponding wave functions onto the Ru orbitals for each layer. 
We used a symmetric slab with 2 surface layers surrounding 3 bulk-like layers [Fig.\,\ref{ldaslab}(a,b)], and the surface structure as determined by LEED  \cite{MooreThesis}. This consists of a 7.46\dg{} rotation and a $\sim$1\% elongation of the surface RuO$_2$ octahedra, as well as slightly modified surface Sr positions. The resulting constant energy contour at 24\,meV binding energy, corresponding to a FS with an averaged electron counting $n_e\!\simeq\!3.91$ to match the $s$-$b$ averaged value experimentally observed by ARPES, is shown in Fig.\,\ref{ldaslab}(c): 2 sets of red FSs pertaining to the 2 outer surface layers, and 3 blue FS sets for the 3 bulk layers; hybridization between layers does occur at some $k$-points, where sheets can be seen to change character. Although somewhat complicated by hybridization gaps appearing at all crossings, and a $\delta_s$ pocket of $d_{x^2-y^2}$ character at $(0,0)$ and ($\pi,\pi)$ which experimentally is found just above $E_F$  \cite{PhysRevB.64.180502}, these calculations show good overall agreement with the phenomenological FS concerning the splitting, topology, and volume of the various bands (Tab.\,\ref{table}); as well as -- and most importantly --  the surface versus bulk-like assignment of all detected features.

To conclude, by means of ARPES and LDA+SO slab calculations we have been able to unravel the surface-to-bulk progression of the electronic structure in \SR{}. We find that a \rtr{} reconstruction of surface and subsurface layers provides a consistent explanation of all detected dispersive features in terms of a progression of electronic states induced by structural instabilities, with no evidence for novel phases driven by topological bulk properties, ferromagnetic ordering, or the interplay between SO and the broken symmetry of the surface. This layer-by-layer approach provides the most detailed information on FS volumes, Fermi velocities, as well as many body renormalizations $v_F^\text{ARPES}/v_F^\text{LDA+SO}$, for both surface and bulk-like bands (Tab.\,\ref{table}). In analogy with the recent work on underdoped cuprates \cite{arXiv1111.2673,NatComm.3.1023}, which found evidence for the enhancement of electronic correlations and ordering tendencies in the surface and subsurface region, our study of \SR{} also highlights the significantly more correlated character of the top surface layer bands, which might thus more strongly benefit from an LDA+SO+U description \cite{PhysRevLett.101.026408}. As a matter of fact, the inclusion of $U\!=\!1$\,eV (not shown) immediately leads to the lifting of the $\delta_s$ FS predicted by LDA+SO, as a result of the overall $U$-driven bandwidth renormalization.

We acknowledge  C. Bergemann,  M.W. Haverkort, R.G. Moore, and G.A. Sawatzky for discussions, D. Wong and P. Dosanjh for technical assistance. This work was supported by the Max Planck -- UBC Centre for Quantum Materials (A.N.), the CRC and NSERC's Steacie Fellowship Programs (A.D.), NSERC, CFI, CIFAR Quantum Materials, and MEXT KAKENHI (No. 22103002).

\bibliography{UnravellingSRO_long}

\end{document}